\documentclass[a4paper,11pt]{article}

\usepackage{jheppub} 

\usepackage[latin3]{inputenc}
\usepackage[makeroom]{cancel}
\usepackage{graphicx}
\usepackage{amsmath}
\usepackage{amsfonts}
\usepackage{amssymb}
\usepackage{color}
\usepackage[export]{adjustbox}
\usepackage{graphicx}
\usepackage{dcolumn}
\usepackage{bm}
\usepackage{simplewick}
\usepackage{array}
\usepackage{appendix}
\usepackage{relsize}
\usepackage{subcaption}

\newcommand{\be}{\begin{equation}}
    \newcommand{\ee}{\end{equation}}
\newcommand{\beq}{\begin{equation}}
    \newcommand{\eeq}{\end{equation}}
\newcommand{\bea}{\begin{eqnarray}}
    \newcommand{\eea}{\end{eqnarray}}

\title{\boldmath Particle dynamics in non-rotating Konoplya and Zhidenko black hole
immersed in an external uniform magnetic field}

\author{Aqeela Razzaq and}
\author{Rehana Rahim }

\affiliation {Department of Mathematics and Statistics, Riphah International
	University, Islamabad, Pakistan}

\emailAdd{aqeelarazzaq45@gmail.com}
\emailAdd{rehana.rahim@riphah.edu.pk}
\abstract{In this paper, we investigate the dynamics of particles in the background of
non-rotating Konoplya and Zhidenko black hole that is immersed in an
external uniform magnetic field. The work involves circular motion of
electric and magnetic particles and particle acceleration. First the motion
of electric charged particles is considered. The effective potential, energy
and angular momentum expressions are obtained along with their graphs. The
analysis of ISCO shows that radii of ISCO decrease with magnetic
interaction parameter. Motion of magnetically charged particles has also
been studied.
\vspace{85 mm} }

\bigskip\bigskip

\notoc

\begin{document}
\maketitle
\flushbottom

\section{Introduction}

Black holes are an interesting and important predictions of Einstein's
theory of general relativity (GR). These are the objects having such an
immense gravitational force that even light cannot escape from them. They
also serve as an excellent laboratory for testing GR in the strong
gravitational field regime. Event horizon around the black holes acts as a
one way membrane from which things do not come out if they enter the event
horizon. Motion of photons and the matter in the close vicinity of a black
hole can help in direct and indirect observation of the event horizon.

The geometric structure of a spacetime can be studied through the analysis
of particle dynamics around a black hole. The motion of charged particles is
affected by the presence of a test uniform magnetic field in the near
vicinity of a black hole. As a black hole hole does not have a magnetic
field, an external magnetic field can be taken into account. Wald gave
solution of the electromagnetic field equations for the Kerr black hole
surrounded by an asymptotically uniform magnetic filed \cite{wald}.
Afterward, many studies have been devoted for the investigation of
electromagnetic fields around black holes surrounded by the external uniform
and dipolar magnetic fields \cite{dm1,dm2,dm3,dm4}. The strength of the
magnetic field is assumed to be weak and particles are taken to be of mass
which is negligible as compared to the black hole's mass.

Currently, GR is the best theory which describes gravity, having passed the
testing with flying colors in the weak gravitational field regime. In GR, Kerr black hole describes the
metric around an astrophysical black hole. Kerr metric contains two parameters, which are mass and spin (and
charge in case of Kerr-Newman black hole). In alternate theories of
gravity, numerous metrics have been developed which contain deviations from
Kerr \cite{jp,cjp1,urk,cpr,ks,kono}. The Kerr metric is obtained when the
deviations vanish. In this paper, we consider the modified Kerr
metric developed in Ref. \cite{kono} by Konoplya and Zhidenko (referred in
this work as KZ black hole). The main aim behind this metric was to see if
the detection of gravitational waves lead to the possibility of modified
theories of gravity \cite{kono1}. Some studies also suggest that a KZ
spacetime might describe a real astrophysical black hole \cite{kono2}.

The paper is arranged as: Section \ref{metric} describes the Konoplya and
Zhidenko black hole. In Section \ref{me}, the magnetic field components are
determined. In Sections \ref{ep} and \ref{mpp}, motion of electric and
magnetic charged particles is discussed, respectively. Center of mass energy
for the collision of two particles is studied in Section \ref{ecm}. The work
has been concluded in the last section.

\section{The Konoplya and Zhidenko black hole}

\label{metric} The rotating Konoplya and Zhidenko black hole metric is given
as \cite{kono}
\begin{align}
ds^{2}&=-\left(1-\frac{2Mr^{2}+\eta }{r\Sigma}\right)dt^{2}+\frac{\Sigma}{
\Delta } dr^{2}+\Sigma d\theta ^{2}{+}\sin ^{2}\theta \left(r^{2}{+}a^{2}{+}
\frac{ (2Mr^{2}{+}\eta )a^{2}\sin ^{2}\theta }{r\Sigma}\right)d\phi ^{2}
\notag \\
& -\frac{ 2(2Mr^{2}+\eta )a\sin ^{2}\theta }{r\Sigma}dtd\phi ,  \label{1}
\end{align}
with
\begin{equation}
\Sigma=r^{2}+a^{2}\cos ^{2}\theta, \quad \Delta =a^{2}+r^{2}-2Mr-\frac{\eta
}{r},  \label{2}
\end{equation}
where $M$ is the mass and $a$ is spin parameter of black hole. Deviations
from Kerr metric are measured by parameter $\eta$. Equation (\ref{1})
becomes Kerr metric when $\eta$ is set to zero. To obtain the non-rotating
form, the case of $a=0$ is considered. This gives
\begin{equation}
ds^{2}=-f(r) dt^{2}+\frac{1}{f(r) } dr^{2}+r ^{2}d\theta ^{2}+r^{2}\sin
^{2}\theta d\phi ^{2},  \label{3}
\end{equation}
with
\begin{equation}
f(r)=\frac{r^3-2Mr^{2}-\eta }{r^{3}}.  \label{4}
\end{equation}
This article deals with the non-rotating form of metric (\ref{1}) shown in Eq. (\ref{3}).

\section{Magnetized Konoplya and Zhidenko black hole}

\label{me} 
\begin{figure}[!htb]
\minipage{0.5\textwidth} \includegraphics[width=2.9 in]{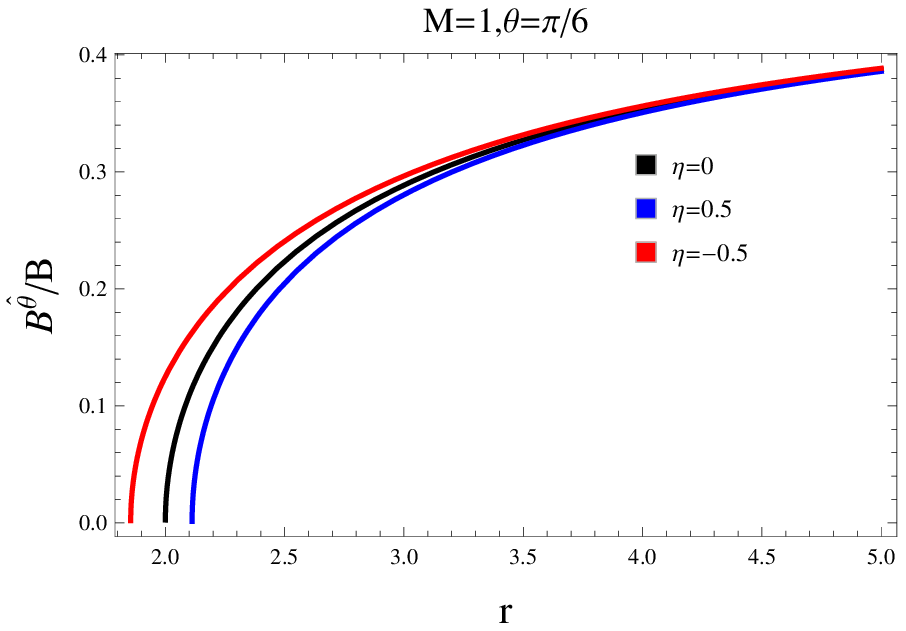}
\endminipage
\hfill \minipage{0.5\textwidth} \includegraphics[width=2.9
	in]{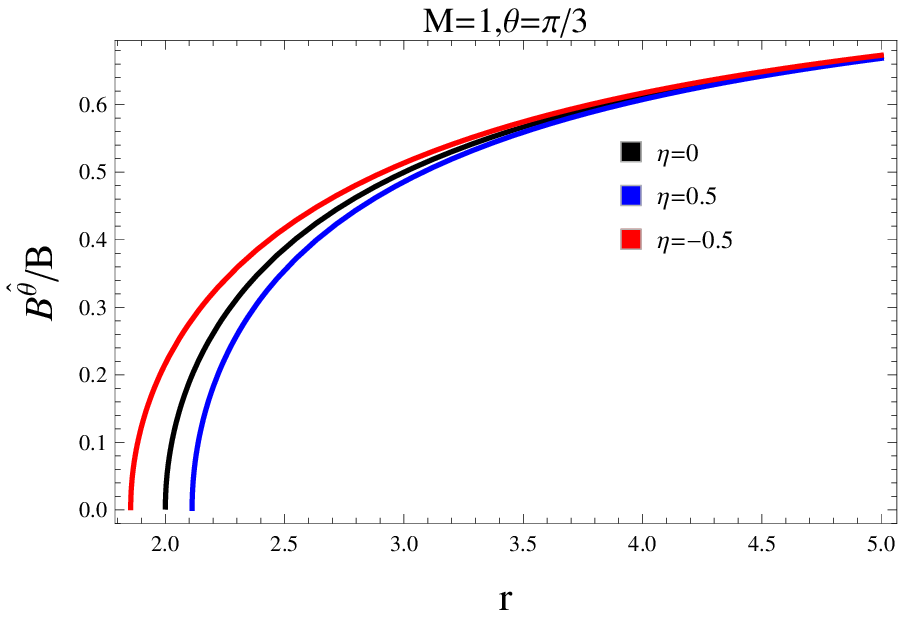} \endminipage
\hfill \minipage{0.5\textwidth} \includegraphics[width=2.9
	in]{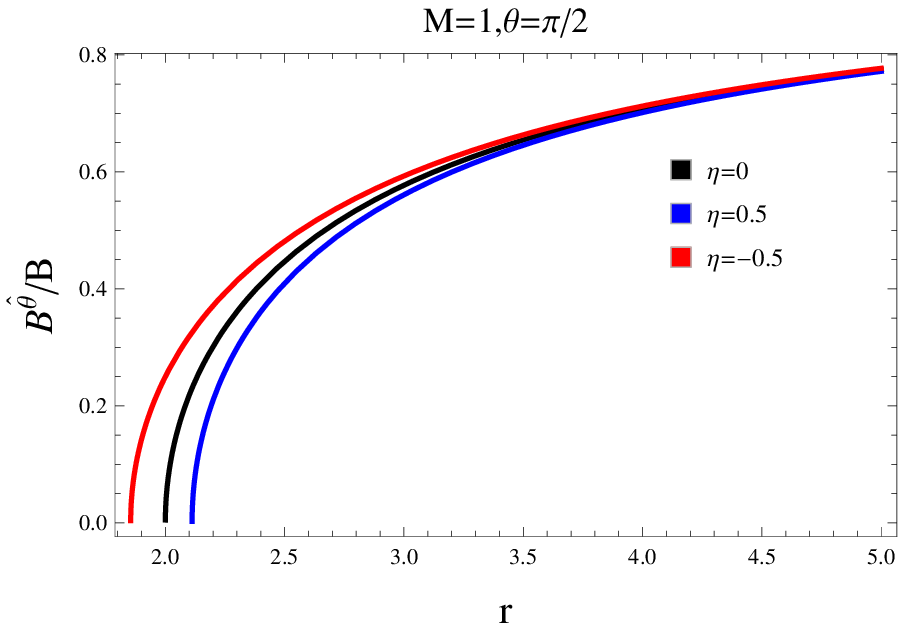} \endminipage
\caption{Radial profile of $B^{\hat{\protect\theta}}$ for some values of
$\eta $ and $\theta $. }
\label{magg1}
\end{figure}

In this section, we consider metric (\ref{3}) surrounded by an
external uniform magnetic field of strength $B$. The magnetic field is taken
to be static, axially symmetric and homogeneous at spatial infinity. It is
also taken to be weak so that it does not effect the spacetime geometry
outside the black hole. Electromagnetic 4-potential determined through Wald
method is \cite{wald}
\begin{equation}
A_{\mu }=\left( 0,0,0,\frac{1}{2}Br^{2}\sin ^{2}\theta \right) .  \label{47}
\end{equation}%
The Maxwell tensor in terms of $A_{\mu }$ is
\begin{equation}
F_{\alpha \beta }=A_{\beta ,\alpha }-A_{\alpha ,\beta },  \label{49}
\end{equation}%
with the components
\begin{equation}
F_{r\phi }=Br\sin ^{2}\theta ,  \label{50}
\end{equation}%
\begin{equation}
F_{\theta \phi }=Br^{2}\sin \theta \cos \theta .
\end{equation}
The orthonormal components of magnetic field with respect to chosen frame
are
\begin{align}
& B^{\hat{r}}=B\cos \theta ,  \label{m1} \\
& B^{\hat{\theta}}=\sqrt{\frac{r^{3}-2Mr^{2}-\eta }{r^{3}}}B\sin \theta .
\label{m2}
\end{align}%
The plot of $B^{\hat{\theta}}$ against various values of $\eta $ and $\theta
$ has been shown in FIG. (\ref{magg1}). From this figure, it is observed
that $B^{\hat{\theta}}$ increase with decreasing value of $\eta $.

\section{The motion of the electric charged particles}

\label{ep}

\label{ep} 
\begin{figure}[tbh]
\minipage{0.5\textwidth} \includegraphics[width=2.7 in]{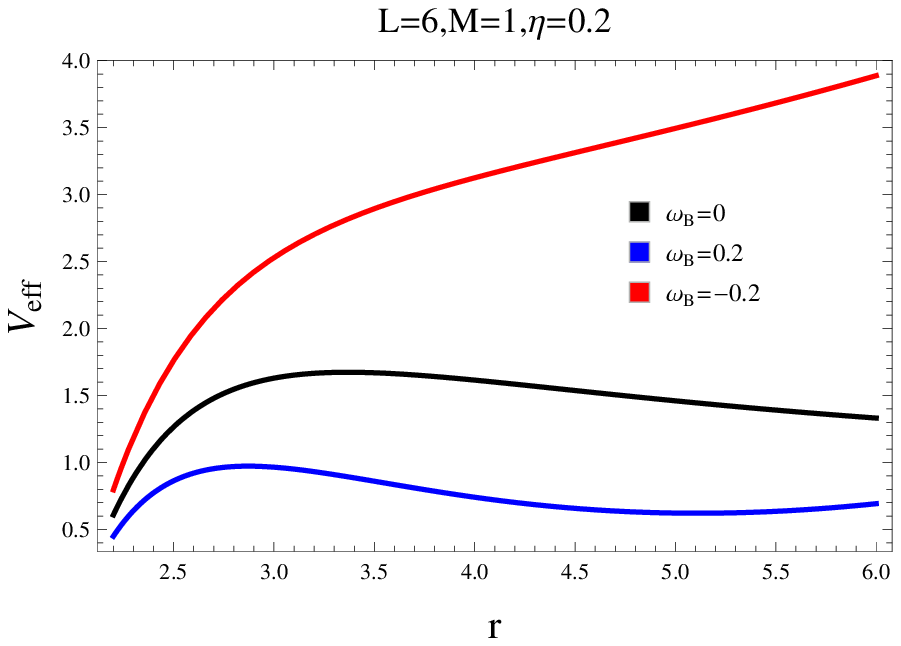}
\endminipage
\hfill \minipage{0.5\textwidth} \includegraphics[width=2.7
	in]{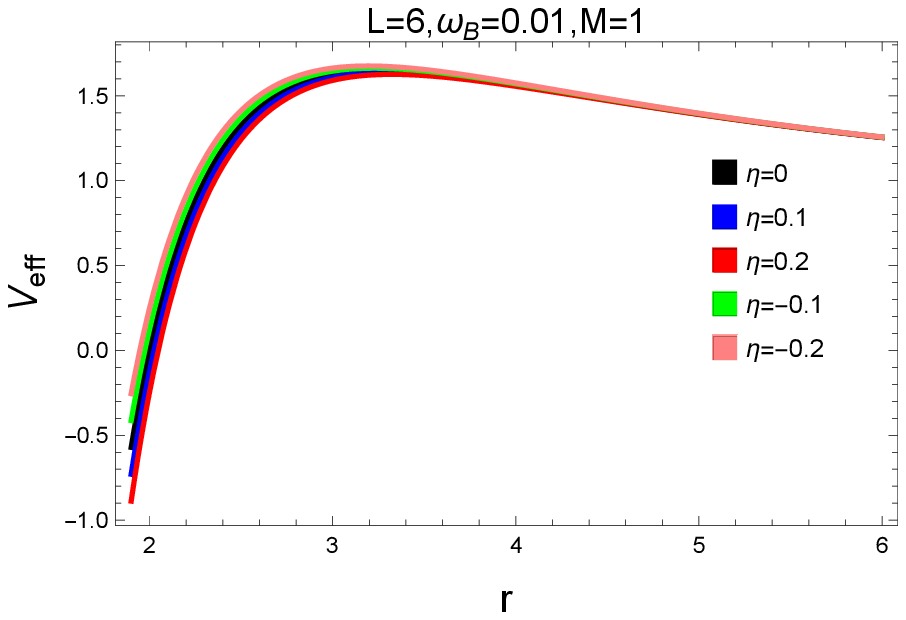} \endminipage
\caption{Radial plot of $V_{eff}$. Graph on the left is shown for some
values of the cyclotron frequency with other parameters being fixed. The
plot in the right panel shown $V_{eff}$ for some values of $\protect\eta$.}
\label{veff}
\end{figure}

This section deals with the circular motion of particles of mass $%
m$ with charge $e$ around the KZ metric, surrounded by an external uniform
magnetic field. Hamilton-Jacobi equation is employed for this purpose and it
is given as
\begin{equation}
g^{\mu \nu }\bigg(\frac{\partial S}{\partial x^{\mu }}-eA_{\mu }\bigg)\bigg(%
\frac{\partial S}{\partial x^{\nu }}-eA_{\nu }\bigg)=-m^{2},  \label{m3}
\end{equation}%
where $S$ is the Hamilton-Jacobi action having the following equation
\begin{equation}
S=-Et+L\phi +S_{r}(r)+S_{\theta}(\theta),  \label{m4}
\end{equation}%
with $E$ being energy and $L$ being angular momentum of the particle,
receptively. The motion takes place on the equatorial plane ($\theta =\pi /2$%
). Equation (\ref{m3}) after putting the values, takes the form
\begin{equation}
\left( \frac{-r^{3}}{r^{3}-2Mr^{2}-\eta }\right) E^{2}+\left( \frac{%
r^{3}-2Mr^{2}-\eta }{r^{3}}\right) g_{rr}^{2}\dot{r}^{2}+\frac{1}{r^{2}}%
(L-eA_{\phi })^{2}=-m^{2}.
\end{equation}%
Further simplification, leads to
\begin{equation}
\dot{r}^{2}=\varepsilon ^{2}-\left( \frac{r^{3}-2Mr^{2}-\eta }{r^{3}}\right)
\left( 1+\left( \frac{\mathit{\mathcal{L}}}{r}\mathit{-\omega }_{B}r\right)
^{2}\right) ,  \label{m5}
\end{equation}%
where $\varepsilon =E/m$, $\mathcal{L}=L/m$ be the energy per unit mass,
angular momentum per unit mass, respectively, and
\begin{equation}
\mathit{\omega }_{B}=\frac{\mathit{eB}}{2m}.  \label{m6}
\end{equation}%
The $\omega _{B}$ is the cyclotron frequency. It accounts for the magnetic
interaction between an electric charge and an external magnetic field.
Equation (\ref{m5}) can also be written as
\begin{equation}
\dot{r}^{2}=\varepsilon ^{2}-V_{eff},  \label{m7}
\end{equation}%
where
\begin{equation}
V_{eff}=\left( \frac{r^{3}-2Mr^{2}-\eta }{r^{3}}\right) \left( 1+\left(
\frac{\mathit{\mathcal{L}}}{r}\mathit{-\omega }_{B}r\right) ^{2}\right) .
\label{m8}
\end{equation}%
The radial plot of $V_{eff}$ has been shown in FIG. (\ref{veff}). The plots
show that if we increase values of $\omega _{B}$ and $\eta$, effective
potential decreases. 
\newline
For circular motion of the particles, one needs the
conditions
\begin{align}
& \dot{r}=0,  \label{ro} \\
& dV_{eff}/dr=0.  \label{vf}
\end{align}%
Equation (\ref{ro}) leads to
\begin{equation}
V_{eff}=\varepsilon ^{2},  \label{m9}
\end{equation}%
while Eq. (\ref{vf}) gives
\begin{align}
V_{eff}^{^{\prime }}& =\frac{1}{r^{6}}\bigg(-2Lr^{2}\omega _{B}\left( 3\eta
+2Mr^{2}\right) +\omega _{B}^{2}\left( -2Mr^{6}+2r^{7}+\eta r^{4}\right)
+L^{2}\left( 5\eta +6Mr^{2}-2r^{3}\right)  \notag \\
& +2Mr^{4}+3\eta r^{2}\bigg)=0.
\end{align}%
This gives angular momentum $\mathcal{L}$ as
\begin{align}
\mathcal{L}& =\frac{1}{2(6Mr^{2}-2r^{3}+5\eta )}\bigg[4Mr^{4}\omega
_{B}+6r^{2}\mathit{\omega }_{B}\eta \pm 2r\big[3\eta \left( 2r^{3}-5\eta
\right) -12M^{2}r^{4}  \notag \\
& +4M(r^{5}-7\eta r^{2})+4r^{2}\omega _{B}^{2}\left( \eta
+2Mr^{2}-r^{3}\right) ^{2}\big]^{1/2}\bigg].
\end{align}%
The energy per unit mass $\varepsilon $ is obtained as
\begin{eqnarray}
{\LARGE \varepsilon }^{2} &=&\left( 1-\frac{2Mr^{2}+\eta }{r^{3}}\right) \\
&&\left( 1+\left( r\omega _{B}-\frac{4Mr^{4}\omega _{B}+6r^{2}\eta \omega
_{B}-\left(
\begin{array}{c}
\left( 4Mr^{4}\omega _{B}+6r^{2}\eta \omega _{B}\right) ^{2}-4\left(
-6Mr^{2}+2r^{3}-5\eta \right) \\
\left( -2Mr^{4}-3r^{2}\eta +2Mr^{6}\omega _{B}^{2}-2r^{7}\omega
_{B}^{2}-r^{4}\eta \omega _{B}^{2}\right)%
\end{array}%
\right) ^{1/2}}{2r\left( 6Mr^{2}-2r^{3}+5\eta \right) }\right) ^{2}\right) .
\notag
\end{eqnarray}%
\begin{figure}[tbh]
\centering
\includegraphics[scale=0.8]{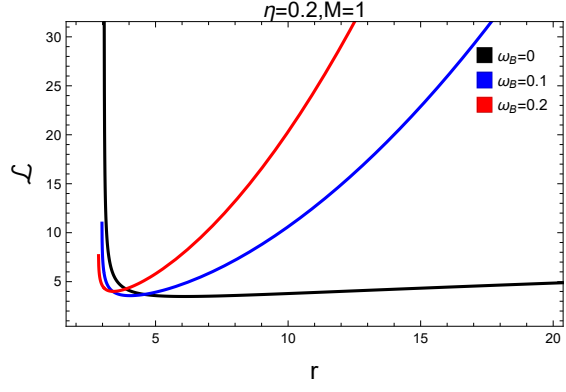} 
\caption{ The graph of $\mathcal{L}$ for some values of $\protect\omega _{B}$
}
\label{l1}
\end{figure}
\begin{figure}[tbh]
\minipage{0.5\textwidth} \includegraphics[width=2.9 in]{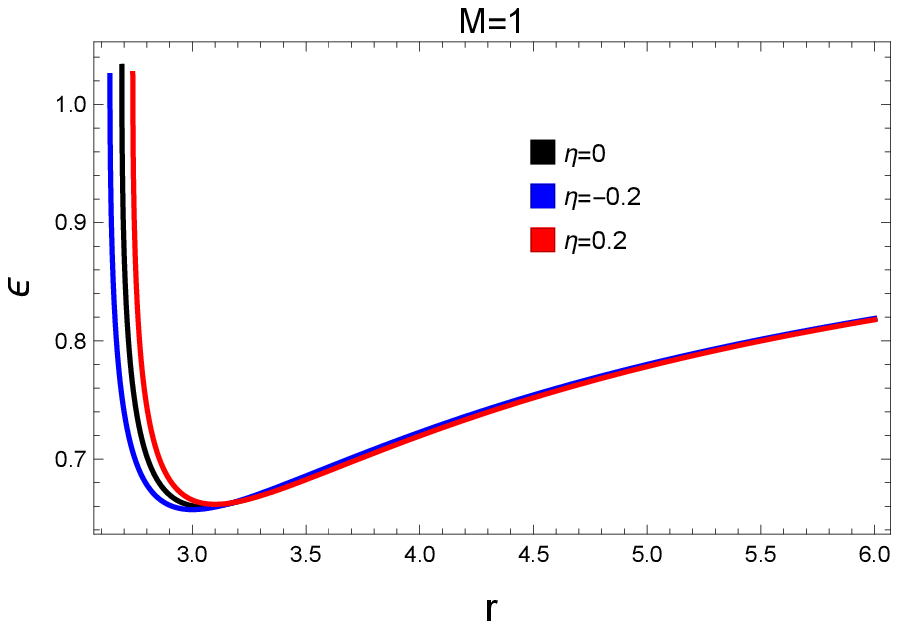}
\endminipage
\hfill \minipage{0.5\textwidth} \includegraphics[width=2.9
	in]{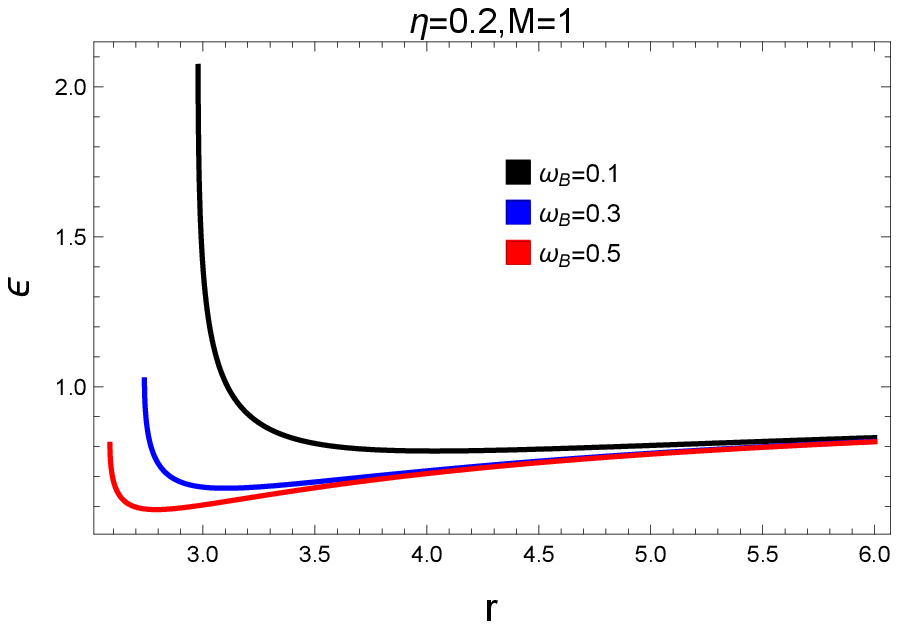} \endminipage
\caption{ Graph of energy of a charged particle. 
 In the left panel
$\protect\omega _{B}=0.3$ with varying values of $\protect\eta $. In the
right panel, graph is shown for some values of $\protect\omega _{B}$ with $%
\protect\eta $ being held fixed.}
\label{ec1}
\end{figure}
The graphical behavior of angular momentum is shown in FIG. 3 which shows
large values of angular momentum due to $\omega _{B}$. Energy is observed to
decrease with $\omega _{B}$ on the right panel of FIG. 4, while, in its left
panel, energy is less than the case of Schwarzschild metric.

\subsection{The inner most stable circular orbits (ISCO)}
To find inner most stable circular orbits or the ISCO, we have $%
d^{2}V_{eff}/dr^{2}=0.$ This gives
\begin{equation*}
\frac{2}{r^{5}\left( 6Mr^{2}-2r^{3}+5\eta \right) ^{2}}\left(
\begin{array}{c}
\left( \left( 6Mr^{2}-2r^{3}+5\eta \right) \left(
\begin{array}{c}
2M\left( 6M-r\right) r^{4}+ \\
r^{2}\left( 20M+3r\right) \eta +15\eta ^{2}%
\end{array}%
\right) \right) \\
+4r^{2}\left( -2Mr^{2}+r^{3}-\eta \right) \left(
\begin{array}{c}
24M^{2}r^{4}-22Mr^{5}+4r^{6}+ \\
40Mr^{2}\eta -23r^{3}\eta +10\eta ^{2}%
\end{array}%
\right) \omega _{B}^{2} \\
-2\left( 2M\left( 6M-r\right) r^{4}+r^{2}\left( 20M+3r\right) \eta +15\eta
^{2}\right) \omega _{B} \\
\times \sqrt{%
\begin{array}{c}
r^{2}\left( 2r^{2}\left( -3M+r\right) -5\eta \right) \left( 2Mr^{2}+3\eta
\right) \\
+4r^{4}\left( 2Mr^{2}-r^{3}+\eta \right) ^{2}\omega _{B}^{2}%
\end{array}%
},%
\end{array}%
\right) \geqslant 0.
\end{equation*}

It is impossible to have exact solution for $r_{isco}$, therefore, it is
obtained numerically. The numerical solution is shown for various values of $%
\omega _{B}$ in Table 1. The table shows decreasing $r_{isco}$ with
increasing magnetic interaction parameter.
\begin{equation}
\begin{tabular}{|l|l|}
\hline
$\omega _{B}$ & $r_{isco\text{ }}$ \\ \hline
0.4 & 4.367406 \\ \hline
0.45 & 4.361155 \\ \hline
0.5 & 4.356620 \\ \hline
0.55 & 4.353231 \\ \hline
0.6 & 4.350632 \\ \hline
0.65 & 4.348598 \\ \hline
0.7 & 4.346976 \\ \hline
\end{tabular}%
\end{equation}

\section{Magnetized Particle Motion}

\label{mpp} 
\begin{figure}[tbh]
\minipage{0.5\textwidth} \includegraphics[width=2.9 in]{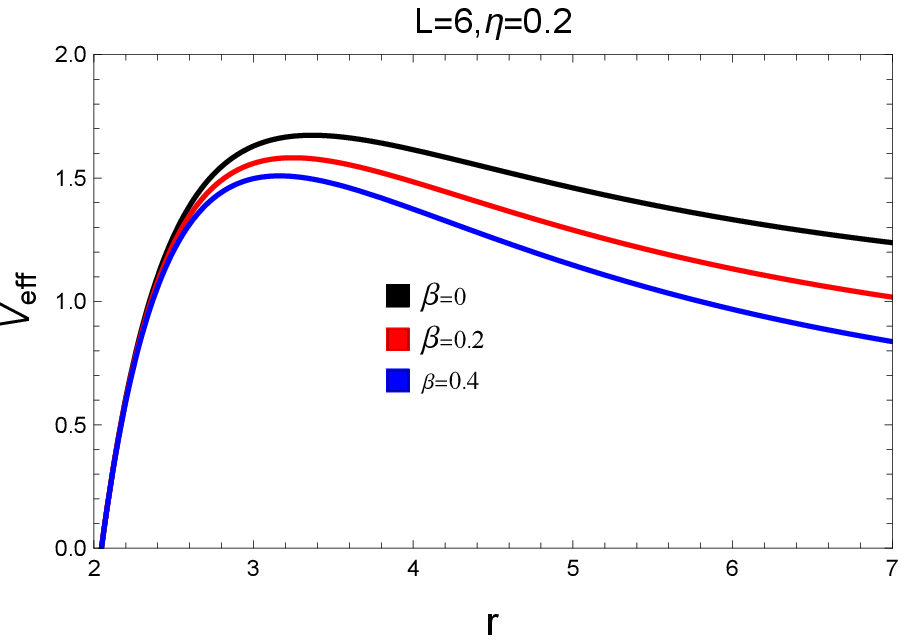}
\endminipage
\hfill \minipage{0.5\textwidth} \includegraphics[width=2.9
	in]{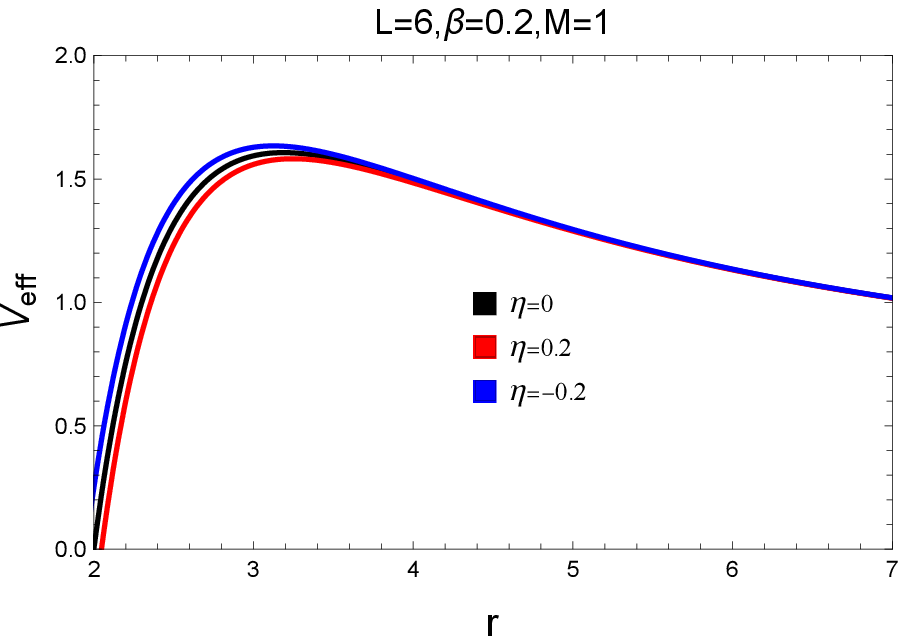} \endminipage
\caption{Graphical representation of $V_{eff}$ of radial motion of
magnetized particles. On the left, graph is drawn for varying $\protect\beta
$. The graph in the right panel has been plotted with varying $\protect\eta $%
. }
\label{vm1}
\end{figure}

\begin{figure}[tbh]
\minipage{0.5\textwidth} \includegraphics[width=2.9 in]{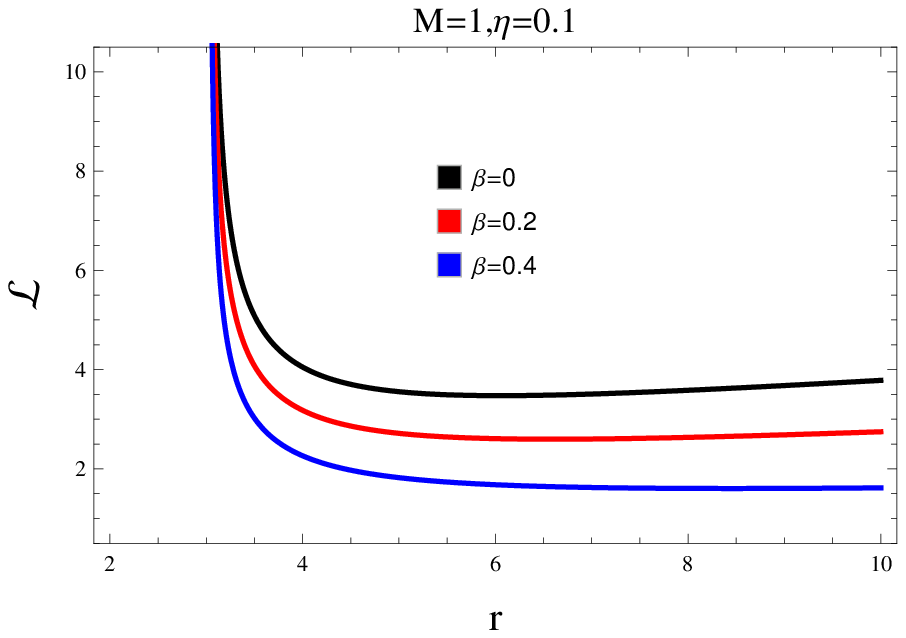}
\endminipage
\hfill \minipage{0.5\textwidth} \includegraphics[width=2.9
	in]{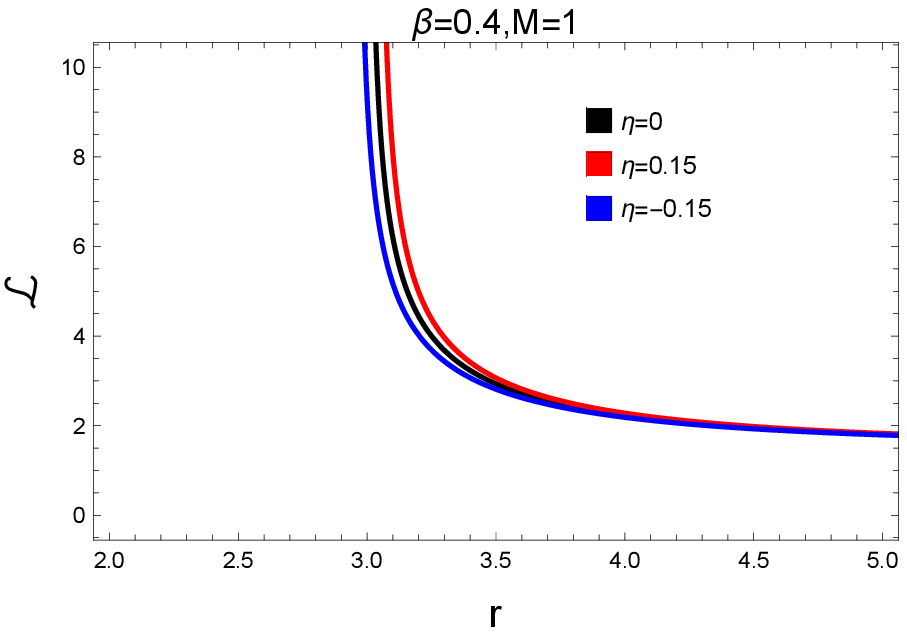} \endminipage
\caption{Graph of $\mathit{\mathcal{L}}$ with varying $\protect\beta $ (left
panel) and $\protect\eta $ (right panel).}
\label{mo}
\end{figure}

\begin{figure}[tbh]
\centering
\includegraphics[scale=0.8]{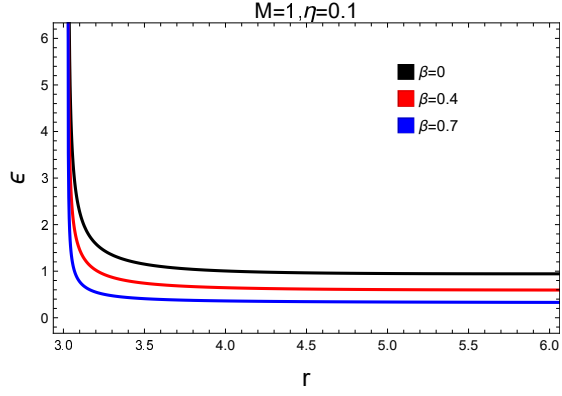} 
\caption{Graphical representation of energy for various values of $\protect%
\beta $.}
\label{en}
\end{figure}

This section deals with the dynamics of magnetized particles
around KZ black hole that is immersed in an external asymptotically uniform
magnetic field. Modified form of Eq. (\ref{m3})
for motion of magnetized particles is
\begin{equation}
g^{\alpha \beta }\frac{\partial S}{\partial x^{\alpha }}\frac{\partial S}{%
\partial x^{\beta }}=-\left( m-\frac{1}{2}D^{\alpha \beta }F_{\alpha \beta
}\right) ^{2},  \label{mpa}
\end{equation}%
with $m$ being particle's mass, $S$ denotes action for magnetized particle
in the curved spacetime background. $D^{\mu \nu }$ represents the
polarization tensor with the form
\begin{equation}
D^{\alpha \beta }=\eta ^{\alpha \beta \mu \nu }u_{\mu }\mu _{\nu },
\label{mp11}
\end{equation}%
and has the following constraint
\begin{equation}
\quad D^{\alpha \beta }u_{\beta }=0.  \label{mp2}
\end{equation}%
Here $\mu _{\nu }$ denotes the 4-velocity of magnetic dipole moment and $%
u_{\nu }$ is the 4-velocity of the particles in an arbitrary observer's rest
frame of reference. The product of $D^{\mu \nu }F_{\mu \nu }$ accounts for
relationship between the external magnetic field and magnetized particles.
In this work, we assume that such an interaction is weak, thus one can
neglect $(D^{\mu \nu }F_{\mu \nu })^{2}.$
The Maxwell tensor can be written as
\begin{equation}
F_{\alpha \beta }=2u_{[\mu }E_{\nu ]}-\eta _{\alpha \beta \mu \nu }B^{\mu
}u^{\nu },  \label{mp3}
\end{equation}%
where $E_{\nu }$ and $B_{\nu }$ are the electric and magnetic field,
respectively, and $\eta _{\alpha \beta \mu \nu }$ is obtained from the
Levi-civita symbol $\epsilon _{\alpha \beta \mu \nu }$ as
\begin{equation}
\eta _{\alpha \beta \mu \nu }=\epsilon _{\alpha \beta \mu \nu }\sqrt{-g},
\label{mp4}
\end{equation}%
with $g$ being the determinant of the metric. Taking into account Eqs. (\ref%
{mp11})-(\ref{mp4}) leads to
\begin{equation}
D^{\alpha \beta }F_{\alpha \beta }=2\mu ^{\alpha }B_{\alpha }=2\mu ^{\hat{%
\alpha}}B_{\hat{\alpha}}=2\mu B\sqrt{f},
\end{equation}%
where $f(r)$ is given in Eq. (\ref{4}). The radial equation of motion is
obtained from Eq. (\ref{mpa}) and is given as 
\begin{equation}
\left( \frac{-r^{3}}{r^{3}-2Mr^{2}-\eta }\right) \varepsilon ^{2}+\left(
\frac{r^{3}}{r^{3}-2Mr^{2}-\eta }\right) \dot{r}^{2}+\frac{1}{r^{2}}\mathit{%
\mathcal{L}}^{2}=-\left( 1-\frac{\mu B\sqrt{f}}{m}\right) ^{2},
\end{equation}%
\begin{equation}
\dot{r}^{2}=\varepsilon ^{2}-\left( \frac{r^{3}-2Mr^{2}-\eta }{r^{3}}\right) %
\left[ (1-\beta \sqrt{f})^{2}+\frac{\mathit{\mathcal{L}}^{2}}{r^{2}}\right] ,\label{32a}
\end{equation}%
where $\beta =\frac{\mu B}{m}$ is called magnetic coupling parameter that
defines electromagnetic interaction between magnetic dipole and external
magnetic field. Equation (\ref{32a}) can be written as
\begin{equation}
\dot{r}^{2}=\varepsilon ^{2}-V_{eff},
\end{equation}%
with $V_{eff}$
\begin{equation}
V_{eff}=\left( \frac{r^{3}-2Mr^{2}-\eta }{r^{3}}\right) \left[ (1-\beta
\sqrt{f})^{2}+\frac{\mathit{\mathcal{L}}^{2}}{r^{2}}\right] .  \label{vm11}
\end{equation}%
FIG. \ref{vm1} shows graph of $V_{eff}$ of Eq. (\ref{vm11}). Both the panels
show decreasing behavior with increasing $\beta $ (left panel) and $\eta $
(right panel).
\newline
To determine the energy and angular momentum, Eqs. (\ref{ro}-\ref{m9}) are
again employed. The derivative of $V_{eff}$ is
\begin{align}
V_{eff}^{^{\prime }}& =\frac{1}{r^{7}}\Big[\mathit{\mathcal{L}}%
^{2}(6Mr^{3}-2r^{4}+5r\eta )-(2Mr^{2}+3\eta )\big(4Mr^{2}\beta ^{2}+2\beta
^{2}\eta   \notag \\
& +r^{3}\big(-1-2\beta ^{2}+3\beta \sqrt{f}\big)\big)\Big]
\end{align}%
This equation leads to angular momentum $\mathit{\mathcal{%
L}}^{2}$ as
\begin{align}
\mathit{\mathcal{L}}^{2}=& \frac{1}{6Mr^{3}-2r^{4}+5r\eta }\Big[%
-2Mr^{5}+8M^{2}r^{4}\beta ^{2}-4Mr^{5}\beta ^{2}-3r^{3}\eta +16Mr^{2}\beta
^{2}\eta -6r^{3}\beta ^{2}\eta   \notag \\
& +6\beta ^{2}\eta ^{2}+6Mr^{2}\beta \sqrt{f}+9r^{3}\beta \eta \sqrt{f}\Big].
\end{align}%
The expression for energy is
\begin{equation}
{\LARGE \varepsilon }^{2}=\frac{(2Mr^{2}-r^{3}+\eta )\left[
\begin{array}{c}
4M^{2}r^{4}\beta ^{2}-\beta ^{2}\eta ^{2}+2r^{6}\left( 1+\beta -2\beta \sqrt{%
f}\right) +r^{3}\eta \left( -2+\beta \left( -\beta +\sqrt{f}\right) \right)
\\
+2Mr^{5}\left( -2+3\beta \left( -\beta +\sqrt{f}\right) \right)
\end{array}%
\right] }{6Mr^{8}-2r^{9}+5r^{6}\eta }.
\end{equation}%
The graphs of angular momentum and energy are shown in FIGs. (\ref{mo}) and (%
\ref{en}), respectively.
The behavior of angular momentum is increasing with increasing values of $%
\eta $ and if we increase values of $\beta $, angular momentum and energy
decrease.
\newline
The ISCO is given by the equation
\begin{equation*}
\frac{1}{r^{7}}\left(
\begin{array}{c}
-\left( 2Mr^{2}+3\eta \right) \left( 8Mr\beta ^{2}+\frac{6Mr^{2}\beta
+9\beta \eta }{2r\sqrt{f}}+3r^{2}\left( -1-2\beta ^{2}+3\beta \sqrt{f}%
\right) \right)  \\
-4Mr\left( 4Mr^{2}\beta ^{2}+2\beta ^{2}\eta +r^{3}\left( -1-2\beta
^{2}+3\beta \sqrt{f}\right) \right)  \\
+\frac{\left( 2Mr^{2}+3\eta \right) \left( 18Mr^{2}-8r^{3}+5\eta \right)
\left( 4Mr^{2}\beta ^{2}+2\beta ^{2}\eta +r^{3}\left( -1-2\beta ^{2}+3\beta
\sqrt{f}\right) \right) }{6Mr^{3}-2r^{4}+5r\eta }%
\end{array}%
\right) \geq 0.
\end{equation*}%
After simplification, one gets
\begin{equation*}
\frac{1}{r^{7}}\left(
\begin{array}{c}
-\left( 2Mr^{2}+3\eta \right) \left( 8Mr\beta ^{2}+\frac{6Mr^{2}\beta
+9\beta \eta }{2r\sqrt{f}}+3r^{2}\zeta \right)  \\
-4Mr\left( 4Mr^{2}\beta ^{2}+2\beta ^{2}\eta +r^{3}\zeta \right)  \\
+\frac{\left( 2Mr^{2}+3\eta \right) \left( 18Mr^{2}-8r^{3}+5\eta \right)
\left( 4Mr^{2}\beta ^{2}+2\beta ^{2}\eta +r^{3}\zeta \right) }{%
6Mr^{3}-2r^{4}+5r\eta }%
\end{array}%
\right) \geq 0.
\end{equation*}%
with
\begin{equation}
\zeta =\left( -1-2\beta ^{2}+3\beta \sqrt{f}\right) .
\end{equation}

\section{Center of mass energy in the equatorial plane}

\label{ecm}

\label{ECM} This section deals with center of mass energy for the collision
of particles. The particles are assumed to be having equal masses and are
assumed to be coming from infinity with the same initial energy $%
E_{1}/m_{1}=E_{2}/m_{2}=1$ but with different angular momenta. The center of
mass energy for the collision of two particles given by Ba\~{n}ados, Silk and
West (BSW) \cite{20a} as
\begin{equation}
{\Large \varepsilon _{cm}=\frac{{E }_{cm}^{2}}{2m_{0}}=1-g_{\mu\nu}v_{1}^{%
\mu}v_{2}^{\nu},}  \label{70}
\end{equation}
where $v_{i}^{\mu} =(\dot{t}_{i},\dot{r}_{i},\dot{\theta}_{i},\dot{\phi}
_{i})$ for $i=1,2$ represent the velocity of the particles.

\subsection{The center of mass energy for the collision of two neutral
particles}

\label{np}

\begin{figure}[tbh]
\minipage{0.5\textwidth} \includegraphics[width=2.9 in]{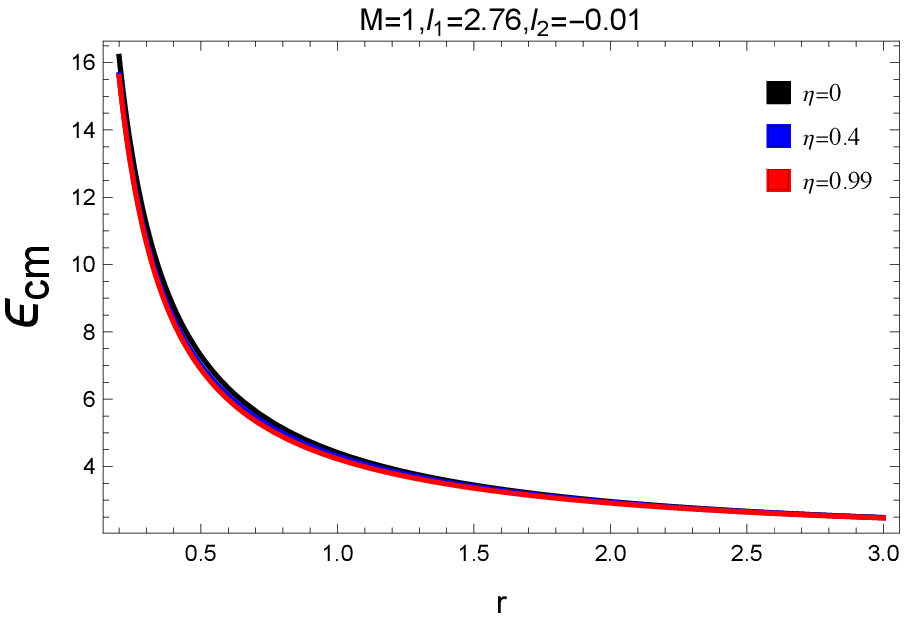} %
\endminipage\hfill \minipage{0.5\textwidth}
\includegraphics[width=2.9
	in]{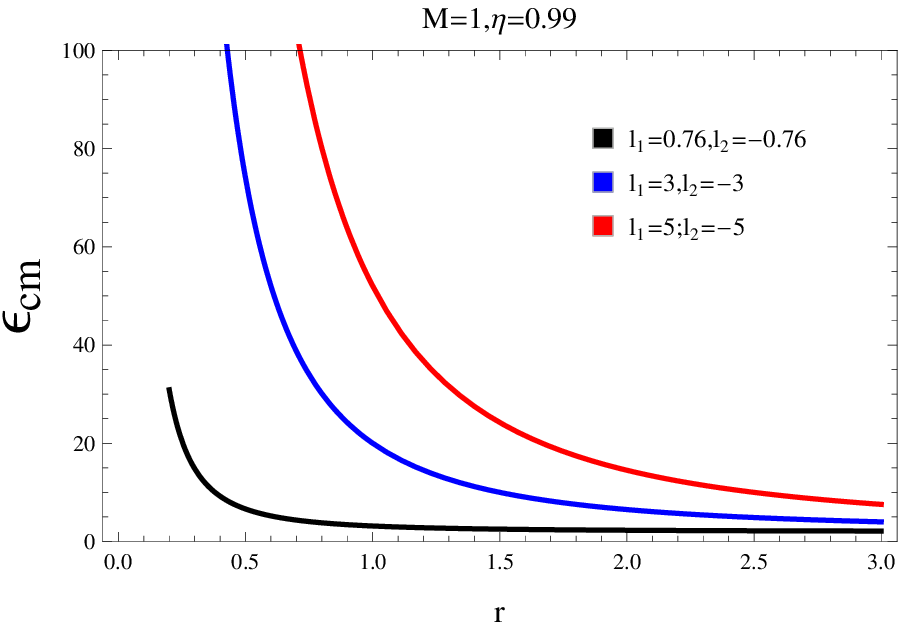} \endminipage
\caption{Graph of ${\protect\Large \protect\varepsilon _{cm}}$ for varying $%
\protect\eta$ (left panel) and angular momentum (right panel). }
\label{ncm}
\end{figure}

In this subsection, collision of two neutral particles having same rest mass
energies will be considered at the equitorial plane. The velocity components
in this case are
\begin{align}
& \dot{t}=\frac{r^{3}}{r^{3}-2Mr^{2}-\eta }, \\
& \dot{\phi}=\frac{l}{r^{2}}, \\
& \dot{r}^{2}=\varepsilon ^{2}-\left( \frac{r^{3}-2Mr^{2}-\eta }{r^{3}}%
\right) \left( 1+\frac{l^{2}}{r^{2}}\right) .
\end{align}%
The center of mass energy is
\begin{align}
\varepsilon _{cm}& =1+\frac{r^{3}}{r^{3}-2Mr^{2}-\eta }-\frac{l_{1}l_{2}}{%
r^{2}}  \notag \\
& -\frac{r^{3}}{r^{3}-2Mr^{2}-\eta }\sqrt{1-\left( \frac{r^{3}-2Mr^{2}-\eta
}{r^{3}}\right) \left( 1+\frac{l_{1}^{2}}{r^{2}}\right) }\sqrt{1-\left(
\frac{r^{3}-2Mr^{2}-\eta }{r^{3}}\right) \left( 1+\frac{l_{2}^{2}}{r^{2}}%
\right) },
\end{align}%
where $l_{1}$ and $l_{2}$, respectively, represent angular momentum of first
and second particle. The radial plot of ${\Large \varepsilon _{cm}}$ is
shown in FIG. \ref{ncm}. In FIG. \ref{ncm} the center of mass energy is
decreasing with increasing values of $\eta $ (left) and increasing with
increasing values of angular momentum.



\subsection{The center of mass energy for the collision of two magnetized
particles}

\label{mp} 
\begin{figure}[tbh]
\minipage{0.5\textwidth} \includegraphics[width=2.9 in]{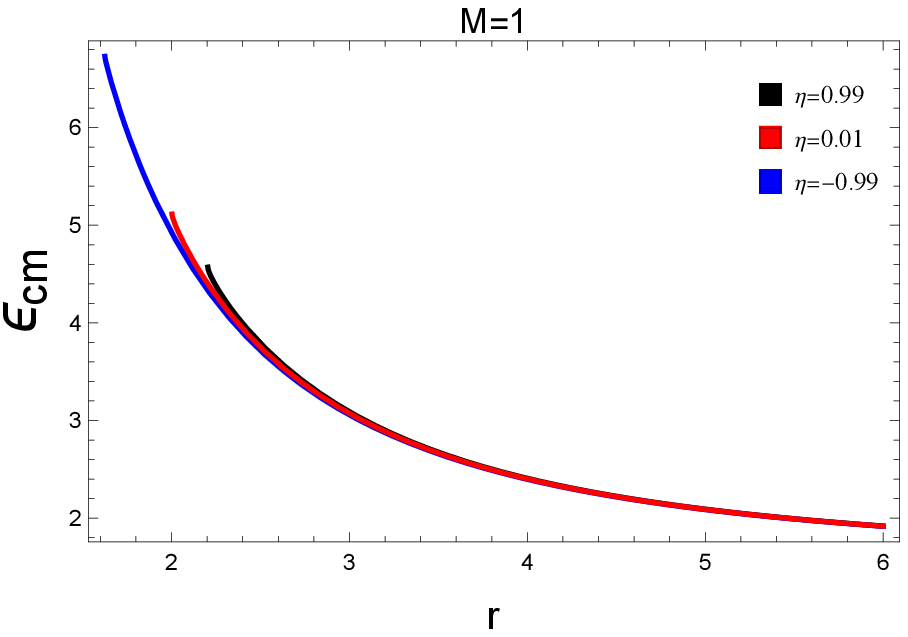} %
\endminipage\hfill \minipage{0.5\textwidth}
\includegraphics[width=2.9
	in]{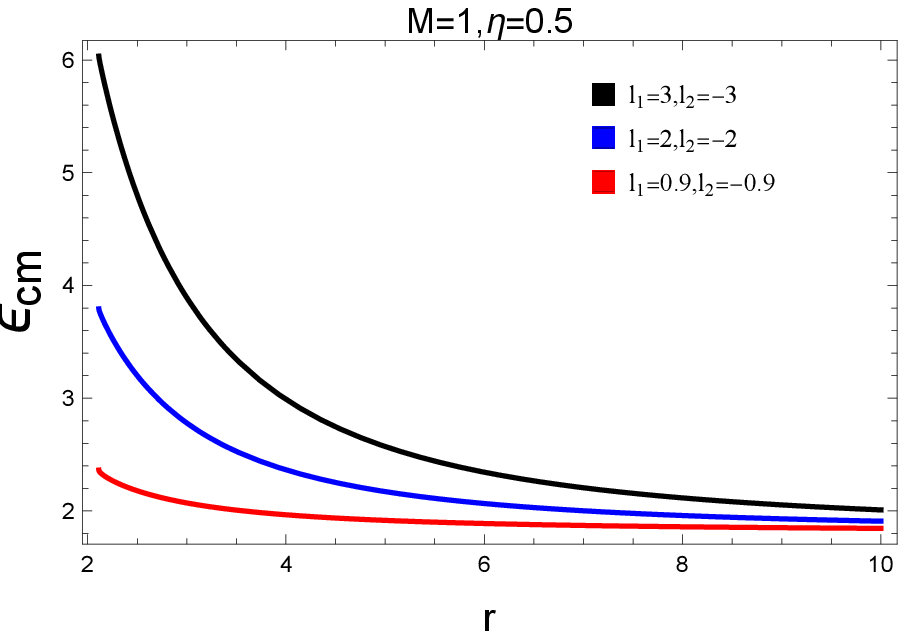} \endminipage
\caption{Center of mass energy for the two magnetized particles collision. On left, $l_{1}=2.5,l_{2}=-2.5,\protect\beta _{1}=0.3=\beta
_{2}.$ On the right, we have taken $\beta _{1}=0.1=\beta _{2}$.}
\label{mm}
\end{figure}

Here, the collision of two magnetized particles will be
considered. The equations for motion in this case are
\begin{align}
& \dot{t}=\frac{r^{3}}{r^{3}-2Mr^{2}-\eta },  \label{mp1} \\
& \dot{\phi}=\frac{l}{r^{2}},  \label{mp2a} \\
& \dot{r}^{2}=\varepsilon ^{2}-\left( \frac{r^{3}-2Mr^{2}-\eta }{r^{3}}%
\right) \left( (1-\beta \sqrt{f})^{2}+\frac{l^{2}}{r^{2}}\right) .
\label{mp3a}
\end{align}%
The center of mass energy in this case is
\begin{equation*}
\begin{array}{c}
{\Large \varepsilon }_{cm}^{2}=1+\frac{r^{3}}{r^{3}-2Mr^{2}-\eta }-\frac{%
l_{1}l_{2}}{r^{2}}-\frac{r^{3}}{r^{3}-2Mr^{2}-\eta }\sqrt{1-\left( \frac{%
r^{3}-2Mr^{2}-\eta }{r^{3}}\right) \left( (1-\beta _{1}\sqrt{f})^{2}+\frac{%
l_{1}^{2}}{r^{2}}\right) }\times \\
\sqrt{1-\left( \frac{r^{3}-2Mr^{2}-\eta }{r^{3}}\right) \left( (1-\beta _{2}%
\sqrt{f})^{2}+\frac{l_{2}^{2}}{r^{2}}\right) }.%
\end{array}%
\end{equation*}
The graphical behavior is shown in FIG. \ref{mm}. In FIG. \ref{mm} if we
decrease values of deformation parameter and angular momentum then the
center of mass energy also decreases.

\subsection{The center of mass energy for the collision of a neutral and a
magnetized particle}

\begin{figure}[tbh]
\centering
\includegraphics[scale=0.8]{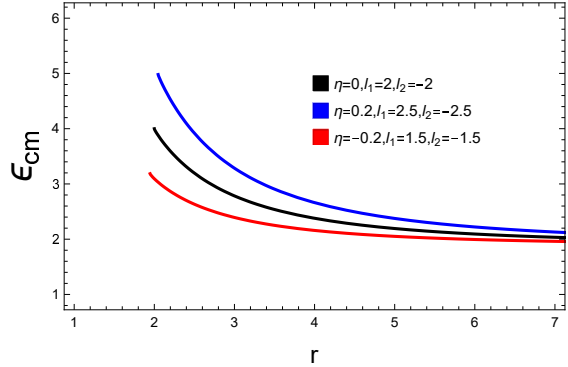} 
\caption{ Radial profile of ${\protect\Large \protect\varepsilon }$ with
varying $\protect\eta $ and $l_{1}$ and $l_{2}$. Here we have taken $M=1,%
\protect\beta _{1}=0.2$. }
\label{mnc}
\end{figure}

In this subsection, collision of a magnetized and neutral particle has been
considered. The equations of motion are given in sections \ref{np} and \ref%
{mp}. The particle 1 is taken to be magnetized and particle 2 is assumed to
be neutral. Using these in center of mass energy expression
\begin{align}
{\Large \varepsilon }_{cm}^{2}=& 1+\frac{r^{3}}{r^{3}-2Mr^{2}-\eta }-\frac{%
l_{1}l_{2}}{r^{2}}-\frac{r^{3}}{r^{3}-2Mr^{2}-\eta }  \notag \\
& \times \sqrt{1-\left( \frac{r^{3}-2Mr^{2}-\eta }{r^{3}}\right) \left(
(1-\beta _{1}\sqrt{f})^{2}+\frac{l_{1}^{2}}{r^{2}}\right) }\sqrt{1-\left(
\frac{r^{3}-2Mr^{2}-\eta }{r^{3}}\right) \left( 1+\frac{l_{2}^{2}}{r^{2}}%
\right) }.
\end{align}%
The radial profile of ${\Large \varepsilon }$ is shown in FIG. \ref{mnc}. In
FIG. \ref{mnc} if we increase values of $\eta $ center of mass energy
increases.

\subsection{The center of mass energy for the collision of a charged and a
magnetized particle}

\begin{figure}[tbh]
\centering
\includegraphics[scale=0.8]{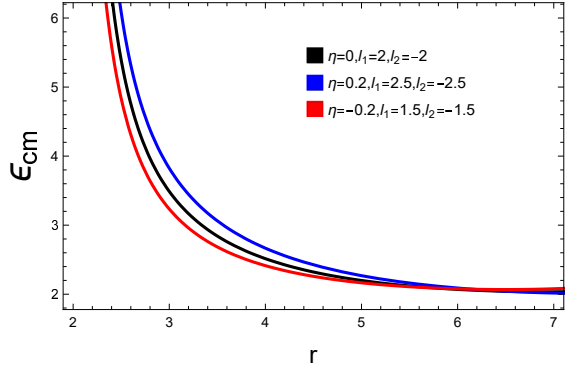} 
\caption{ Radial profile of ${\protect\Large \protect\varepsilon }$ with
varying $\protect\eta $ and $l_{1}$ and $l_{2}$. Here we have taken $M=1,%
\protect\omega _{B}=0.1,\protect\beta =0.1$. }
\label{ecmp}
\end{figure}

Here, collision of a magnetized and a charged particle has been considered.
The particle 1 is taken to be charged and particle 2 is assumed to be
magnetized. Using these in center of mass energy expression, we obtain

\begin{align}
{\Large \varepsilon }_{cm}^{2}& =1+\frac{r^{3}}{r^{3}-2Mr^{2}-\eta }-\left(
\frac{l_{1}}{r^{2}}-\omega _{B}\right) l_{2}-\frac{r^{3}}{r^{3}-2Mr^{2}-\eta
}  \notag \\
& \times \sqrt{1-\left( \frac{r^{3}-2Mr^{2}-\eta }{r^{3}}\right) \left[
1+\left( \frac{l_{1}}{r}-\omega _{B}r\right) ^{2}\right] \text{ }}\sqrt{%
1-\left( \frac{r^{3}-2Mr^{2}-\eta }{r^{3}}\right) \left( (1{-}\beta\sqrt{f}%
)^{2}{+}\frac{l_{2}^{2}}{r^{2}}\right) }.
\end{align}%
The center of mass energy is shown in FIG. \ref{ecmp}. In FIG. \ref{ecmp} if
we increase values of $\eta $ center of mass energy increases.

\section{Summary and conclusion}

The Kerr black hole solution is an axisymmetric, stationary and vacuum
solution of the Einstein theory of general relativity. All the astrophysical
black holes are expected to be described by the Kerr metric. There has been
a lot of interest in modified Kerr black hole solutions. Such solutions
contain parameters which account for possible deviations from Kerr, which is
obtained when deviations are set to zero. One such rotating metric has been
developed by Konoplya and Zhidenko \cite{kono}, whose non-rotating case has
been discussed in the present article. 
In this work, dynamics of charged and magnetized particles (in the
equatorial plane) have been discussed in the background of non-rotating
Konoplya-Zhidenko metric immersed in an external magnetic field. \newline
First, the effective potential, angular momentum and energy for the circular
motion of charged test particles have been studied for the dependence on
deformation parameter $\eta $ and cyclotron frequency $\omega _{B}$.
In FIG. \ref{veff}, effective potential has been plotted for varying
$\omega _{B}$ (left panel) and $\eta $ (right panel). Both panels show
deceasing trend with increasing the respective parameter. This can be
explained as the decrease in the least distance between the charged
particles and the black hole with increase of $\omega _{B}$ and $\eta $. In
FIGs. \ref{l1} and \ref{ec1}, radial plots of angular momentum and energy
are shown. If we take double derivative of effective potential it give us
value of ISCO and it is not possible to find its analytical solution so we
just solve it numerically to check its behavior. The table 1. shows the ISCO
values at different values of $\omega _{B}$ with $\eta$ fixed. A decreasing
trend is observed for this table. 
Decreasing the ISCO radius is very important because the gravitational
potential near the central object can accelerate particles to high energies.

Section \ref{mpp} contains dynamics of magnetized particle in the background
of non-rotating KZ black hole. First the effective has been shown in FIG. %
\ref{vm1}. The figure shows the same decreasing trend for $\beta$ and $\eta$
as in the charged particles case. Next, angular momentum, energy and ISCO
have been studied. In the last section, center of mass energy was studied.
The exact expression were shown along with the graphical behavior in each
case.


\begin{thebibliography}{99}
\bibitem{wald} R. M. Wald, Phys. Rev. D \textbf{10} 1680 (1974).

\bibitem{dm1} A. A. Abdujabbarov, B. J. Ahmedov and V. G. Kagramanova, Gen.
Relativ. Gravit. \textbf{40} 2515 (2008).

\bibitem{dm2} A. N. Aliev and D. V. Gal'tsov, Soviet Phys. Uspekhi \textbf{%
32 } 75 (1989).

\bibitem{dm3} A. N. Aliev, D. V. Galtsov and V. I. Petukhov, Astrophys. Sp.
Sci. \textbf{124} 137 (1986).

\bibitem{dm4} T. Oteev, A. Abdujabbarov, Z. Stuchlik and B. Ahmedov,
Astrophys. Sp. Sci. \textbf{361} 269 (2016).

\bibitem{jp} T. Johannsen and D. Psaltis, Phys. Rev. D \textbf{83} 124015
(2011).

\bibitem{cjp1} R. Rahim and K. Saifullah, Ann. Phys. \textbf{405} 220 (2019).

\bibitem{urk} U. A. Gillani, R. Rahim and K. Saifullah, Astropart. Phys.
\textbf{138} 102684 (2022).

\bibitem{cpr} R. Rahim and K. Saifullah, IJMPD Doi:
10.1142/S0218271821501236 (2021).

\bibitem{ks} K. Glampedakis and S. Babak, Class. Quantum Grav. \textbf{23}
4167 (2006).

\bibitem{kono} R. Konoplya and A. Zhidenko, Phys. Lett. B \textbf{\ 756 }350
(2016).

\bibitem{kono1} F. Long. S. Chen, S. Wang and J. Jing, A. Zhidenko, Nucl.
Phys. B. \textbf{\ 926}83 (2018).

\bibitem{kono2} C. Bambi, S. Nampalliwar, Europhys. Lett. \textbf{116} 30006
(2016); Y. Ni, J. Jiang, C. Bambi, J. Cosmol. Astropart. Phys. \textbf{9} 14
(2016).

\bibitem{20a} M. Ba\~{n}ados, J. Silk and S. M. West, Phys. Rev. Lett. \textbf{%
103} (2009) 111102.
\end{thebibliography}
\end{document}